\documentclass[11pt]{article}
\usepackage[usenames,dvipsnames]{color}
\usepackage{hyperref}

\begin{document}
\title
{Effect of a cosmological constant on propagation of vacuum polarized photons in stationary spacetimes}
\author
{Sourav Bhattacharya{\footnote{souravbhatta@physics.uoc.gr}}
\\
Institute for Theoretical and Computational Physics, Department of Physics\\ University of Crete, 700 13 Heraklion, Greece\\
}

\maketitle
%%%%%%%%%%%%%%%%%%%%%%%%%%%%%%%%%%%%%%%%%%%%%%%%%%%%%%%%%%%%%%%
\abstract
Consideration of vacuum polarization in quantum electrodynamics may affect the momentum dispersion relation for photons for a non-trivial background, due to appearance of curvature dependent terms in the effective action. We investigate the effect of a positive cosmological constant $\Lambda$ on this at one loop order for stationary $\Lambda$-vacuum spacetimes. To the best of our knowledge, so far it has been shown that $\Lambda$
affects the propagation in a time dependent black hole spacetime. Here we consider the static de Sitter cosmic string and the Kerr-de Sitter spacetime to show that there can be some non-vanishing effect due to $\Lambda$ for physical polarizations. Consistency of these results with the polarization sum rule is discussed.  
\vskip .5cm

{\bf Keywords:} {Vacuum polarization, curved spacetime, cosmological constant
  
%\maketitle

\vskip 1cm

%%%%%%%%%%%%%%%%%%%%%%%%%%%%%%%%%%%%%%%%%%%%%%%%              SECTION 1
\maketitle
\section{Introduction}
%\vskip 0.1cm
%\noindent

%\maketitle
%\noindent\textbf{A. Introduction}
%\vskip 0.1cm
\noindent
If we consider quantum electrodynamics in curved spacetime, the effective action for the photon contains from the loop integrals some finite part which depends on both the field strength as well as the background curvature. Such finite parts originate from the curvature dependence of the propagator in curved spacetime. In that case the photon equation of motion gets modified and consequently the dispersion relation for null geodesics may no longer hold due to the gravitational tidal forces, as was first shown in~\cite{Drummond:1979pp}.  

%In particular, when one takes quantum electrodynamics (QED) interactions in a general non-singular curved spacetime, the effect of vacuum polarization modifies the photon propagator in a way mentioned above and one finds super/subluminal photons for physical polarizations~\cite{Drummond:1979pp}. Analogous result was discussed for massless spin-$\frac12$ fields~\cite{Ohkuwa:1980jx}.

%\noindent 
Since then there has been a lot of attention given to this exotic effect~\cite{Hollowood:2012as},~\cite{Daniels:1993yi}-\cite{Hollowood:2011yh}.  Such effect may also be present for massless spin-1/2 fields~\cite{Ohkuwa:1980jx}, gravitons~\cite{Stanley:2011kz}, and even in flat spacetimes with non-trivial vacuum, e.g.~\cite{Latorre:1994cv, Dittrich:1998fy}.
% with non-trivial topology or boundary~\cite{Barton:1989dq, Scharnhorst:1990sr}, or in presence of strong background classical electromagnetic field which breaks the symmetry of the Minkowski vacuum~\cite{Dittrich:1998fy}.
 
%\noindent 
The velocity shifts of photons were investigated in various black hole backgrounds~\cite{Daniels:1993yi, Daniels:1995yw, Cho:1997vg, Cai:1998ij}. 
%In~\cite{Daniels:1995yw}, a general formula for velocity shift of photons with the averaged sum over physical polarizations in general curved spacetimes was given. 
%In~\cite{Shore:1995fz}, it was shown that the velocity shift on a black hole horizon vanishes for velocity direction orthogonal to the horizon.
 For various aspects associated with this phenomenon including the preservation of unitarity and causality, we refer our reader to~\cite{Shore:1995fz,Hollowood:2012as, Shore:2003zc}.

%\noindent 
The chief concern of this work is to investigate the role of a positive cosmological constant $\Lambda$ in this phenomenon.  
The physical motivation of this study  comes from the observation of accelerated expansion of our universe~\cite{Riess:1998cb, Perlmutter:1998np}. It was shown in~\cite{Drummond:1979pp} that there is no modification of photon's
speed in any maximally symmetric space like de Sitter. In~\cite{Cai:1998ij} various stationary and time dependent black holes with $\Lambda$  were considered and was shown that although the velocity of the vacuum polarized photons gets modified, $\Lambda$  contributes to this only in the time dependent case. Therefore, it is interesting to see whether we may find some non-trivial effect due to $\Lambda$ in some physically interesting stationary spacetimes.  

 %\noindent 
In particular, we shall consider below two $\Lambda>0$ spacetimes -- the Linet-Tian cosmic string~\cite{Tian:1986zz, Bhattacharya:2008fu} and the Kerr-de Sitter~\cite{Carter:1968ks} to show that we may have non-vanishing effects due to $\Lambda$, for physical or transverse polarizations. %It is clear that the Kerr-de Sitter universe can be a very good physical model for a rotating stellar object in our universe. 

%\noindent 
For the cosmic string spacetime in particular, the false vacuum of the scalar field inside the string core makes the effective value of the cosmological constant higher and hence for such spacetimes this effect may be considerable inside the string core.   

%We wish to emphasize here that all these calculations are valid in the weak curvature limit, where one retains only terms linear in curvature in the so called Gilkey-Seeley-DeWitt expansion~\cite{Parker} of the effective action. Thus as pointed out in~\cite{Drummond:1979pp},
%the application of these results in early universe, where curvature is strong, does not seem to be very meaningful. Secondly, these calculations uses some local normal coordinate neighborhood around a point, and one ignores quantum fluctuations due to particle creation or vacuum ambiguity. For this reason this calculations does not seem to be application in the global cosmological scenario.
%In fact, this kind of effect seems to be verifiable via some local birefringence effect~\cite{Drummond:1979pp}, which is of course, far beyond the reach of current day observation.    

%In the following we shall briefly outline the basic scheme and the calculation of the photon's velocity shifts.
%We shall use mostly negative signature for the metric $(+,-,-,-)$ and will set $G=1$ throughout.

%\vskip .2cm

%%%%%%%%%%%%%%%%%%%%%%%%%%%%%%%%%%%%%%%%%%%%%%%%%%%%%%    SECTION 2
\noindent
\section{Calculation of the velocity shifts}
%\vskip 0.2cm
%%%%%%%%%%%%%%%%%%%%%%%%%%%%%%%
\noindent 
The basic calculational scheme uses the 1-loop corrected effective equation of motion for photons in a weakly curved space for QED~\cite{Drummond:1979pp, Shore:1995fz, Cai:1998ij},
%\newpage
%
%\begin{widetext}
\begin{eqnarray}
\nabla_aF^{ab}+\frac{2}{m_{\rm e}^2}\nabla_a\left[ 2 a R F^{ab}
+b\left(R^a{}_cF^{cb}-R^b{}_cF^{c a}\right)\right]
\nonumber\\ +\frac{4c}{m_{\rm e}^2}\nabla_a\left(R^{ab}{}_{cd}F^{cd}  \right)=0,
\label{lp1}
\end{eqnarray}
%\end{widetext}
%
where $m_{\rm e}$ is the electron mass and $a=-\frac{5\alpha}{720 \pi}$, $b=\frac{26\alpha}{720\pi}$ and $c=-\frac{2\alpha}{720\pi}$,
and $\alpha$ is the fine structure constant.
%One then decomposes the field strength tensor into a fixed background 
%$\overline{F}^{ab}$ and a propagating part $f^{ab}$, and employs the usual eikonal ansatz $f_{ab}=\widetilde{f}_{ab}e^{i\chi(x)}$. The amplitude $\widetilde{f}_{ab}$ is assumed to be `slowly' varying with respect to the variation of the phase. One also assumes that the variation of the phase is much larger than the variation of the spacetime curvature. The wave vector is given by the normal to the constant phase hypersurfaces : $k_a=\nabla_a\chi$. Putting these all in
Using eikonal approximation Eq.~(\ref{lp1}) can be cast in the leading order~\cite{Drummond:1979pp},
%
%\begin{widetext}
\begin{eqnarray}
k^2 a_b + \frac{2b}{m_{\rm e}^2}R_{ac}\left(k^ak^c a_b-k^aa^ck_b\right) + \frac{8 c}{m_{\rm e}^2}R_{abcd}k^ak^ca^d=0,%\nonumber\\
\label{lp2}
\end{eqnarray}
%\end{widetext}
%
where $k^a$ is the photon momentum, $k^2=k_ak^a$ and $a_b$ is the polarization vector, $a\cdot k=0$. We put above $R_{ab}=\Lambda g_{ab}$ for $\Lambda$-vacuum Einstein spaces. The above equation is the leading departure from the null geodesic approximation. 
 We may note here that if we assume the spacetime to be maximally symmetric de Sitter, then the Riemann tensor gets decomposed into terms proportional to $\Lambda$ and $g_{ab}$ and accordingly we get $k^2=0$. 

%\noindent 
Precisely, our goal here is to provide examples of non-maximally symmetric stationary $\Lambda$-vacuum Einstein spaces for which not only $k^2\neq 0$, but also $\Lambda$ has non-vanishing contributions into this, by solving Eq.~(\ref{lp2}) for physical or transverse polarizations.

\subsection{The Linet-Tian-cosmic string spacetime} 

 Let us start with the static cylindrical Linet-Tian or Nielsen-Olesen cosmic string spacetimes~\cite{Tian:1986zz, Bhattacharya:2008fu}, the exterior of which is given by 
%\newpage
%
%\begin{widetext}
\begin{eqnarray}
ds^2=\cos^{\frac43}\frac{\rho\sqrt{3\Lambda} }{2}\left(dt^2-dz^2\right)-d\rho^2-\frac{4\delta^2}{3\Lambda}\sin^2\frac{\rho\sqrt{3\Lambda} }{2}\cos^{-\frac23}\frac{\rho\sqrt{3\Lambda} }{2}d\phi^2,
\label{lp3}
\end{eqnarray}
%\end{widetext}
%
%\newpage
where $\delta$ is a constant representing the conical singularity. The inside core metric for the Nielsen-Olesen configuration near the axis looks formally the same as above, with $\delta=1$, and $\Lambda$ replaced with an effective cosmological constant $\Lambda'=\Lambda+2\pi\lambda \eta^4$~\cite{Bhattacharya:2008fu}, where $\lambda$
and $\eta$ are respectively the coupling constant for quartic self interaction and the expectation value of the complex scalar field in its false vacuum. The above spacetime has a curvature singularity at $\rho=\frac{\pi}{\sqrt{3\Lambda}}$~\cite{Tian:1986zz}. Since Eq.~(\ref{lp2}) is valid only in the weak curvature limit, the following calculations will not hold near that singularity. Clearly, the above metric globally cannot represent a physical spacetime itself, and it has to be embedded within some other spacetime to get a reasonable global scenario. In any case, the Linet-Tian spacetime can be a very good realistic model for spacetimes with a positive cosmological constant in the vicinity of a cylindrical mass distribution of compact cross section.   

%\noindent 
We choose orthonormal basis : $\omega^0=\sqrt{g_{tt}}dt,~\omega^1=\sqrt{-g_{\phi\phi}}d\phi,~\omega^2= d\rho,~\omega^3=\sqrt{-g_{zz}}dz$. Following e.g.~\cite{Cai:1998ij}, we next define tensors : $U^{\mu\nu}_{ab}=\delta^{\mu}_{[a}\delta^{\nu}_{b]}$. With the help of these tensors, we express the components of the Riemann tensor as  (see e.g. Chapter 6 of~\cite{Chandrasekhar:1985kt}, for expressions in generic stationary axisymmetric spacetimes),  
%
%\begin{widetext}
%\begin{eqnarray}
%R_{1010}=R_{2020}=-R_{1313}=-R_{2323}=\frac{\Lambda}{2}\left(1+\frac13\tan^2\left( \frac{\rho\sqrt{3\Lambda} }{2}\right)\right),~R_{1212}=-R_{3030}=\frac{\Lambda}{3}\tan^2\left( \frac{\rho\sqrt{3\Lambda} }{2}\right).
%\label{lp4}
%\end{eqnarray}
%\end{widetext}
%
%Following e.g.~\cite{Cai:1998ij}, we define tensors as  $U^{\mu\nu}_{ab}=\delta^{\mu}_{[a}\delta^{\nu}_{b]}$, with $\mu$ and $\nu$ correspond to the orthonormal frame. With this, the Riemann tensor components of Eq.~(\ref{lp4}) can be written as
%
%\begin{widetext}
\begin{eqnarray}
R_{abcd}=E\left(U_{ab}^{12}U_{cd}^{12}-U_{ab}^{30}U_{cd}^{30} \right)+F\left(U^{10}_{ab}U^{10}_{cd}+U^{20}_{ab}U^{20}_{cd}-U^{23}_{ab}U^{23}_{cd}-U^{31}_{ab}U^{31}_{cd} \right),%s\nonumber\\
\label{lp5}
\end{eqnarray}
%\end{widetext}
%
where $E(\rho)=\frac{\Lambda}{3}\tan^2\left( \frac{\rho\sqrt{3\Lambda} }{2}\right)$, and $F(\rho)=\frac{\Lambda}{2}\left(1+\frac13\tan^2\left(\frac{\rho\sqrt{3\Lambda} }{2}\right)\right)$, so that $R_{1010}=F(\rho)$, and so on.

%\noindent 
Since the conical singularity term $\delta$ in Eq.~(\ref{lp3}) does not contribute to the curvature, it is clear that the above expressions remain formally unchanged near the axis of the string, with $\Lambda$ replaced with $\Lambda'=\Lambda+2\pi\lambda \eta^4$. 

%\noindent 
Also, we define the individual momenta through $U_{ab}^{\mu\nu}$ : $k^bU_{ab}^{\mu\nu}$,
%
%\begin{widetext}
\begin{eqnarray}
l_a=\left(k_0\delta^1_a+k_1\delta^0_a\right),~m_a=\left(k_0\delta^2_a+k_2\delta^0_a\right),~n_a=\left(k_0\delta^3_a+k_3\delta^0_a\right),\nonumber\\ p_a=\left(k_1\delta^2_a-k_2\delta^1_a\right),~q_a=\left(k_2\delta^3_a-k_3\delta^2_a\right),~r_a=\left(k_3\delta^1_a-k_1\delta^3_a\right).%\nonumber\\
\label{lp6}
\end{eqnarray}
%\end{widetext}
%
We contract Eq.~(\ref{lp2}) with a vector field $v^b$, where $v^b$ at a time corresponds to any one of the above momenta and
using Eq.s~(\ref{lp5}), (\ref{lp6}) we find	
%
%\begin{widetext}
\begin{eqnarray}
[1+ \frac{2b\Lambda}{m_{\rm e}^2}] k^2(v\cdot a) - \frac{8 c\Lambda E}{m_{\rm e}^2}\left[(v\cdot p)(a\cdot p) -(v \cdot n)(a\cdot n)\right]\nonumber\\- \frac{8 c\Lambda F }{m_{\rm e}^2}\left[(v\cdot l)(a\cdot l)+(v\cdot m)(a\cdot m)-(v\cdot q)(a\cdot q)-(v\cdot r)(a\cdot r)\right]    =0.	
\label{lp7}
\end{eqnarray}
%\end{widetext}
%
%We shall contract the above equation with a vector field $v^b$ which corresponds to any one of the direction of propagation in Eq.s~(\ref{lp6}), and find the dispersion relation obeyed by $k^a$. We shall only consider physical or transverse polarizations.
We start by considering the radial photon motion, $k\equiv \{k_0,0,k_2,0\}$ in Eq.s~(\ref{lp6}) and $k^2=k_0^2-k_2^2=-m^2$. The two physical polarization vectors for radial photons correspond to the azimuthal and axial directions ($a\sim l,n$).

%\noindent 
For radial photons we have $l\equiv\{0,k_0,0,0\}$ in Eq.~(\ref{lp6}). First we take $v=l$ in Eq.~(\ref{lp7}). Also we have in this case $p=-\frac{k_2}{k_0}l$ and $l\cdot m=l\cdot n=l\cdot q=l\cdot r=0$. Putting these all in together, we get 
%
%\begin{widetext}te
\begin{eqnarray}
k^2(a\cdot l) \left(1+\frac{2b\Lambda}{m_{\rm e}^2}\right)  +\frac{8 c}{m_{\rm e}^2} \left[Ek_2^2 +Fk_0^2  \right] (a\cdot l)  =0,
\label{lp8}
\end{eqnarray}
%\end{widetext}
%
with no mixing term $a\cdot n$ with the other physical polarization vector.
Putting in the expressions of $E$ and $F$ (see after Eq.~(\ref{lp5})), we get, since $a\cdot l\neq 0$, 
%
%%\begin{widetext}
\begin{eqnarray}
\left\vert{\frac{k_0}{k_2}}\right\vert_{\phi} 
=\left(1-\frac{2c\Lambda}{m_{\rm e}^2}\sec^2 \left(\frac{\rho\sqrt{3\Lambda}}{2}\right) \right),
\label{lp9}
\end{eqnarray}
%%\end{widetext}
%
which clearly shows that the null geodesic dispersion relation is violated due to $\Lambda$ via quantum effect, and since $c=-\frac{2\alpha}{720\pi}$, it is superluminal.

%\noindent 
In order to get the equation corresponding to the other polarization vector ($a\sim n$), we take $v=n$ in Eq.~(\ref{lp7}). 
We note from (\ref{lp6}) that $q=\frac{k_2}{k_0}n $ and $n\cdot l=n\cdot m =n\cdot p=n\cdot r=0$. Putting these all in together, we solve Eq.~(\ref{lp7}) as earlier to get 
\begin{eqnarray}
\left\vert{\frac{k_0}{k_2}}\right\vert_{z} 
=\left(1+\frac{2c\Lambda}{m_{\rm e}^2}\sec^2 \left(\frac{\rho\sqrt{3\Lambda}}{2}\right) \right),
\label{lp10}
\end{eqnarray}
which is subluminal.

%\noindent 
Next, let us consider the azimuthal or orbital photons, $k\equiv \{k_0,k_1,0,0\}$, and $k^2=k_0^2-k_1^2=-l^2$, from Eq.s~(\ref{lp6}). The two physical polarization states correspond to the radial and the axial directions, $a\sim m, n$.

%\noindent 
We first take $v=m$ in Eq.~(\ref{lp7}). We have $m\cdot l=m\cdot n=m\cdot q=m\cdot r=0$, and $p=\frac{k_1}{k_0}m $. Then proceding as earlier, from Eq.~(\ref{lp7}) to find 
\begin{eqnarray}
\left\vert{\frac{k_0}{k_1}}\right\vert_{\rho} 
=\left(1-\frac{2c\Lambda}{m_{\rm e}^2}\sec^2 \left(\frac{\rho\sqrt{3\Lambda}}{2}\right) \right).
\label{lp11}
\end{eqnarray}
Likewise, taking $v=n$ and $r=-\frac{k_1}{k_0}n$ and noting that the inner products of $n$ with the vectors $(l,m,p,q)$ vanish, Eq.~(\ref{lp7}) gives
\begin{eqnarray}
\left\vert{\frac{k_0}{k_1}}\right\vert_{z} 
=\left(1+\frac{2c\Lambda}{m_{\rm e}^2}\sec^2 \left(\frac{\rho\sqrt{3\Lambda}}{2}\right) \right).
\label{lp12}
\end{eqnarray}
Finally, we come to the photons moving in the axial or $z$-directions, $k\equiv\{k_0,0,0,k_3\}$, and $k^2=k_0^2-k_3^2=-n^2$, in Eq.(\ref{lp6}). For physical polarization along the radial direction, we have $a\cdot m\neq 0$ and $q=-\frac{k_3}{k_0}m$. 
Then choosing $v=m$ in Eq.~(\ref{lp7}), and noting $m\cdot p=m\cdot n=m\cdot l=m\cdot r=0$, we proceed as earlier to get
%
%\begin{eqnarray}
%\left[1+\frac{2b\Lambda}{m_{\rm e}^2}\right]k^2-\frac{4c\Lambda}{m_{\rm e}^2}\left(1+\frac13\tan^2\left(\frac{\rho\sqrt{3\Lambda}}{2}\right)\right)\left(-k_0^2+k_3^2\right)=0,\nonumber\\
%\label{lp12'}
%\end{eqnarray}
%
$k^2=0$, i.e. the null geodesic dispersion relation. For polarization along $\phi$, we choose $v=l$, also giving $k^2=0$.
This fact can be understood as the maximal symmetry of the `$t-z$' plane of the Linet-Tian spacetime (\ref{lp3}).

%\noindent 
For a Nielsen-Olesen cosmic string spacetime, as we mentioned earlier, sufficiently near the axis, $\Lambda$ can be replaced with $\Lambda'=\Lambda+2\pi \lambda\eta^4$, which is due to the false vacuum of the complex scalar field. Then sufficiently
inside the core, the nontrivial results of Eq.s~(\ref{lp9})-(\ref{lp12}) correspond to replacing $\Lambda$ with $\Lambda'$.

\subsection{The Kerr-de Sitter spacetime} 

Let us now come to the Kerr-de Sitter spacetime~\cite{Carter:1968ks}, which reads in the Boyer-Lindquist coordinates,
%
%\begin{widetext}
\begin{eqnarray}
ds^2=-\Sigma\left(\frac{dr^2}{\Delta_r}+\frac{d\theta^2}{R}\right) -\frac{R\sin^2\theta}{\Sigma}\left(\frac{a dt-(r^2+a^2) d\phi}{1+\frac{\Lambda a^2}{3}}\right)^2+
\frac{\Delta_r}{\Sigma}\left( \frac{dt-a\sin^2\theta d\phi}{1+\frac{\Lambda a^2}{3}} \right)^2,
\label{lp13}
\end{eqnarray}
%\end{widetext}
%
where, $\Sigma=r^2+a^2\cos^2\theta$, $R=1+\frac{\Lambda a^2\cos^2\theta}{3}$ and $\Delta_r=(r^2+a^2)(1-\frac{\Lambda r^2}{3})-2Mr$. We shall work in a region far away from the Schwarzschild radius, where we only retain terms linear in the rotation parameter, $a$. Also, owing to the observed tiny value of $\Lambda\sim 10^{-52}{\rm m^{-2}}$, we always have 
$M\sqrt{\Lambda}\ll 1$, even with large masses such as $M\sim 10^{15}M_{\odot}$. Then, even for $a\sim M$, 
we can safely ignore $\Lambda a^2$ terms for practical purposes. This offers a much simplified form of the above metric,
 %
%%\begin{widetext}
\begin{eqnarray}
ds^2\approx fdt^2- f^{-1}dr^2-r^2d\Omega^2+2\omega r^2\sin^2\theta dtd\phi +{\cal{O}}(a^2),%\nonumber\\
\label{lp14}
\end{eqnarray}
%%\end{widetext}
%
where $f(r)=\left(1-\frac{2M}{r}-\frac{\Lambda r^2}{3}\right)$, and $\omega(r)= a\left(\frac{2M}{r^3}+\frac{\Lambda}{3}\right)$.
This can be thought of as the leading axisymmetric deformation of the Schwarzschild-de Sitter spacetime due to rotation. 
Since the diagonal components of the metric represent norms (i.e. scalars) of various coordinate vector fields, they must be even in the rotation parameter, because otherwise they would depend on the sign of $a$, i.e. the direction of rotation of the source and thus we have the above simplification in the leading order. We also note that the effect of $\Lambda$ should be more as we move away from the gravitating source. In this case, keeping only terms linear in the rotation parameter seems a reasonable approximation, as far as we are interested in the qualitative new effects due to $\Lambda$. On the other hand, if we assume that $a\ll M$, Eq.~(\ref{lp14}) holds throughout.

 %\noindent 
We shall see below that such breaking of the spherical symmetry leads to a non-vanishing contribution of $\Lambda$ in photon propagation.

%\noindent 
Following~\cite{Chandrasekhar:1985kt}, we define orthonormal basis $\omega^0=\sqrt{f}dt$, $\omega^1=r\sin\theta(d\phi-\omega dt)$, $\omega^2=f^{-\frac12}dr$ and $\omega^3=rd\theta$, and compute the components of the Riemann tensor up to terms linear in $a$,  
 %
%\begin{widetext}
\begin{eqnarray}
-R_{0101}=-R_{0303}=R_{1212}=R_{2323}=\frac{M}{r^3}-\frac{\Lambda}{3},~R_{0202}=-R_{1313}=\frac{2M}{r^3}+\frac{\Lambda}{3},\nonumber\\
R_{1303}=-R_{1202}=\frac{3Ma\sin\theta}{r^4}\sqrt{f},~
R_{1230}=R_{1023}=-\frac12 R_{1302}=-\frac{3Ma\cos\theta}{r^4}.
\label{lp15}
\end{eqnarray}
%\end{widetext}
%
 %We denote the above independent functions by $A$, $B$ and $C$
%respectively, and as earlier define tensors $U_{ab}^{\mu\nu}=\delta^{\mu}_{[a}\delta^{\nu}_{b]}$, to find up to terms linear in $a$,
Let us first consider the photon motion on the equatorial plane, $\theta=\pi/2$. We express the non-vanishing curvature components in (\ref{lp15}) as earlier as
%
%\begin{widetext}
\begin{eqnarray}
R_{abcd}=A\left(U^{12}_{ab}U^{12}_{cd}+U^{23}_{ab}U^{23}_{cd}-U^{10}_{ab}U^{10}_{cd}-U^{30}_{ab}U^{30}_{cd}\right)+B\left(U^{20}_{ab}U^{20}_{cd}-U^{31}_{ab}U^{31}_{cd}\right)\nonumber\\+C\left(U^{31}_{ab}U^{30}_{cd}+U^{30}_{ab}U^{31}_{cd}+ U^{12}_{ab}U^{20}_{cd}+U^{20}_{ab}U^{21}_{cd}\right),
\label{lp16}
\end{eqnarray}
%\end{widetext}
%
where $A=\frac{M}{r^3}-\frac{\Lambda}{3}$, $B=\frac{2M}{r^3}+\frac{\Lambda}{3}$ and $C=\frac{3Ma\sqrt{f} }{r^4}$.
We substitute Eq.~(\ref{lp16}) into Eq.~(\ref{lp2}) and contract as earlier with $v^b$ to get
%
%\begin{widetext}
\begin{eqnarray}
k^2(1+\frac{2b\Lambda}{m_{\rm e}^2})(a\cdot v)+\frac{8cA}{m_{\rm e}^2}\left[ (p\cdot v)(p\cdot a)+(q\cdot v)(q\cdot a)-(l\cdot v)(l\cdot a)-(n\cdot v)(n\cdot a)\right]\nonumber\\ + \frac{8cB}{m_{\rm e}^2}\left[(m\cdot v)(m\cdot a)-(r\cdot v)(r\cdot a)\right] +
\frac{8cC}{m_{\rm e}^2}\left[(r\cdot v)(n\cdot a)+(r\cdot a)(n\cdot v) +(p\cdot v)(m\cdot a) +(p\cdot a)(m\cdot v)\right]=0, 
\label{lp17}
\end{eqnarray}
%\end{widetext}
%
where the vectors $l,m,n$ etc. are defined in Eq.~(\ref{lp6}).

%\noindent 
Let us start by considering the radial photons, $k\equiv \{k_0,0,k_2,0\}$, and $k^2=-m^2$ in Eq.~(\ref{lp6}). One of the two physical polarizations is along the orthonormal basis $\omega^1$, defined above Eq.~(\ref{lp15}). In this case $a\cdot l\neq 0$. The other physical polarization is along the polar direction, $a\sim n$.

%\noindent 
First we take $v=l$ in Eq.~(\ref{lp17}). Also, for radial photons we have from Eq.~(\ref{lp6}),
$p=-\frac{k_2}{k_0}l$, and $l\cdot q=l\cdot n=l\cdot m=l\cdot r=0 $.  Then it turns out that
%
%%\begin{widetext}
\begin{eqnarray}
k^2\left(1+\frac{2b\Lambda}{m_{\rm e}^2}\right) +\frac{8cA}{m_{\rm e}^2}\left[(p\cdot l)(a\cdot p)-(l\cdot l)(a\cdot l) \right]=0.
\label{lp17'}
\end{eqnarray}
%%\end{widetext}
%
We have for radial photons, $l^2=-k_0^2$, and $(p\cdot l)(p\cdot a)= -k_2^2(a\cdot l)$. Putting this back into Eq.~(\ref{lp17'}), we get $k^2=0$.  

%\noindent 
Likewise, we take $v=n$ in Eq.~(\ref{lp17}) next. We also have $q=\frac{k_2}{k_0}n$ in Eq.~(\ref{lp6}). Procceding as above we get $k^2=0$, as well. Thus for radial photons there is no velocity shifts for physical polarizations.

%\noindent 
We note that retaining generic rotation in the Kerr case gives a velocity shift of at least ${\cal{O}}(a^2)$~\cite{Daniels:1995yw}. Since we are working only with terms linear in the rotation parameter, we have $k^2=0$,
analogous to the static spacetimes~\cite{Cai:1998ij}.

%It is now easy to find that the only non-vanishing term within the square bracket of Eq.~(\ref{lp17}) is $A\left[(q\cdot n)(a\cdot q)-(n\cdot n)(a\cdot n)\right]$.
 %We have $n\cdot n=-k_0^2$, and $(q\cdot n)(a\cdot q)=-k_2^2(a\cdot n)$. This gives $k^2=0$. 
%\noindent 
Let us now consider the orbital photons.
We have, $k\equiv\{k_0,k_1,0,0\}$, $k^2=-l^2$ in Eq.~(\ref{lp6}). The two physical polarization vectors correspond to radial and polar directions, $a\sim m, n$.

%\noindent 
Taking $v=m$ in Eq.~(\ref{lp17}), and $p=\frac{k_1}{k_0}m$ (follows from (\ref{lp6}), for orbital photons), using $m\cdot m=-k_0^2$ and $p\cdot m=-k_0k_1 $, Eq.~(\ref{lp17}) reduces to 
%
%%\begin{widetext}
\begin{eqnarray}
k^2\left(1+\frac{2b\Lambda}{m_{\rm e}^2}\right)-\frac{8c}{m_{\rm e}^2}\left[Ak_1^2+Bk_0^2+2Ck_0k_1\right]=0. 
\label{lp18}
\end{eqnarray}
%%\end{widetext}
%
Since we are working at one-loop order, whenever the momenta are multiplied with the coefficients $b$ or $c$,
we can take the null geodesic dispersion relation. This means for the last term in the above equation, we can take $k_0=\pm k_1$,
where $+(-)$ sign corresponds to prograde (retrograde) orbits respectively.

%\noindent 
Then we find using the explicit expressions for $A$, $B$ and $C$ (see below Eq.~(\ref{lp16})), 
%
%\begin{widetext}
\begin{eqnarray}
\left\vert\frac{k_0}{k_1}\right\vert_r=1+\frac{12cM}{m_{\rm e}^2 r^3}\left[1\pm\frac{2a}{r}\left(1-\frac{2M}{r}-\frac{\Lambda r^2}{3}\right)^{\frac12}\right],
\label{lp19}
\end{eqnarray}
%%\end{widetext}
%
where $+(-)$ sign corresponds to prograde (retrograde) orbits respectively.

%\noindent 
Similarly for the other physical polarization, setting $v=n$  in Eq.~(\ref{lp17}), and using $r=-\frac{k_1}{k_0}n$ (follows from (\ref{lp6}), for this case) yields 
%
%%\begin{widetext}
\begin{eqnarray}
\left\vert\frac{k_0}{k_1}\right\vert_{\theta}=1-\frac{12cM}{m_{\rm e}^2 r^3}\left[1\pm\frac{2a}{r}\left(1-\frac{2M}{r}-\frac{\Lambda r^2}{3}\right)^{\frac12}\right].%\nonumber\\
\label{lp20}
\end{eqnarray}
%%\end{widetext}
%
%Finally, we come to photons moving along the polar angle, $k\equiv \{k_0,0,0,k_3\}$, and $k^2=-n^2$. We also have in Eq.~(\ref{lp6}),
%$q=-\frac{k_3}{k_0}m$ and $r=\frac{k_3}{k_0}l$. 

%For polarization along the radial direction, we have $a\cdot m\neq 0$, and we take $v=m$ in Eq.~(\ref{lp17}). Likewise, for polarization along azimuthal direction, we set $v=l$ in Eq.~(\ref{lp17}). Then proceeding as earlier we find
%Evaluating various inner products using Eq.~(\ref{lp6}), we find
%
%\begin{eqnarray}
%\left\vert\frac{k_0}{k_3}\right\vert_{r,\phi}=1\pm\frac{12cM}{m_{\rm e}^2 r^3}
%\label{lp21}
%\end{eqnarray}
%

%\begin{eqnarray}
%\left\vert\frac{k_0}{k_3}\right\vert_{\phi}=1-\frac{12cM}{m_{\rm e}^2 r^3}.
%\label{lp21}
%\%end{eqnarray}
%
This completes the calculation part for velocity shifts on the equatorial plane. 

%\noindent 
We note that if we set $M=0$ in Eq.s~(\ref{lp19}), (\ref{lp20}), we recover the vanishing result of the empty de Sitter space~\cite{Drummond:1979pp}. 
%If we set $\Lambda=0$, we recover the result of the Schwarzschild spacetime~\cite{Drummond:1979pp}.

%\noindent 
Let us now check the consistency of our results with the sum rule over the physical polarizations. This states that, the averaged sum of the photon's velocity shift for physical polarization equals $-\frac{\left(4b+8c\right)}{m_{\rm e}^2}R_{ab}k^ak^b$~\cite{Shore:1995fz}. We put $R_{ab}=\Lambda g_{ab}$. At one loop order we may take $k^2=0$ on the right hand side. This means on the average there is no velocity shift for photons in $\Lambda$-vacuum spacetimes. This corresponds to the earlier identically vanishing results~\cite{Drummond:1979pp, Cai:1998ij}. For our present case any of our non-vanishing results comes like $1\pm \epsilon$ for any set of two physical polarizations (Eq.s~(\ref{lp9})-(\ref{lp12}), (\ref{lp19}),
(\ref{lp20})). Hence our result is consistent with the polarization sum rule. 

%\noindent 
It is clear that for {\it any} other $\Lambda$-vacuum spaces, the velocity shift, if it exists, should always be of the form $1\pm\epsilon$, for any set of two physical polarizations.

 %\noindent 
Before we end, we shall briefly discuss now about the extension of the above calculations in the off-equatorial plane.
In that case we retain all the curvature components of (\ref{lp15}). Due to this, two different physical polarizations mix with each other, unlike all the previous cases. For example, the equation for the azimuthal photons corresponding to the physical polarizations $(a\sim m,n)$ become
%
%\begin{widetext}
\begin{eqnarray}
k^2(1+\frac{2b\Lambda}{m_{\rm e}^2})(a\cdot m)-\frac{8c}{m_{\rm e}^2}[Ak_1^2+Bk_0^2+2C_1k_0k_1](a\cdot m)-\frac{24c C_2 k_0k_1}{m_{\rm e}^2}(a\cdot n)=0,\nonumber\\
k^2(1+\frac{2b\Lambda}{m_{\rm e}^2})(a\cdot n)+\frac{8c}{m_{\rm e}^2}[Ak_1^2+Bk_0^2+2C_1k_0k_1](a\cdot n)-\frac{24c C_2 k_0k_1}{m_{\rm e}^2}(a\cdot m)=0,
\label{lp21}
\end{eqnarray}
%\end{widetext}
%
where $C_1=\frac{3Ma\sqrt{f}\sin\theta }{r^4}$, and $C_2=-\frac{3Ma\cos\theta }{r^4}$, and $A$ and $B$ are the same as earlier. 
In this case we need to solve the determinantal equation to get the modified dispersion relation. 
It is easy to check that the additional terms coming from the $C_2$ functions do not make any ${\cal{O}}(a)$-contribution.
And also, in Eq.s~(\ref{lp19}), (\ref{lp20}), the factor $2a$ is now replaced with $2a\sin\theta$. This simplification certainly appears due to keeping terms linear in the rotation parameter. 
The $\sin\theta$ term
represents the fact that the effect of rotation is vanishing along the axis, $\theta=0,\pi$ and maximum on the equatorial plane, $\theta=\pi/2$.

%\vskip .2cm
%%%%%%%%%%%%%%%%%%%%%%%%%%%%%%%%%%%%%%%%%%%%%%%%%%%%%%            SECTION 3
%\noindent
\section{Summary and outlook}
%\vskip 0.1cm
%%%%%%%%%%%%%%%%%%%%%%%%%%%%%%%
%In this work we have investigated the effect of a positive cosmological constant on the propagation of the 1-loop vacuum polarized photons. The chief motivation of this study is the recent discovery of accelerated expansion of our universe~\cite{Riess:1998cb,Perlmutter:1998np}. For empty maximally symmetric de Sitter space or static black hole spacetimes with $\Lambda$
%with different topologies, or even for stationary axisymmetric BTZ black hole, it is known that cosmological constant does not affect the photon propagation for any physical polarization~\cite{Cai:1998ij}. Our goal was to find physically relevant stationary spacetimes
%in which $\Lambda$ has indeed some non-vanishing effect. 

Let us summarize the results now.

%\noindent 
We have considered two physically interesting stationary $\Lambda$-vacuum spacetimes -- the de Sitter cylindrical cosmic string and the Kerr-de Sitter. For the cosmic string spacetime, we have found shift for radial and orbital photons (Eq.s~(\ref{lp9})-(\ref{lp12})), but no shift for axial photons. The vanishing of the velocity shift along the axis can be understood as the Lorentz symmetry in the `$t-z$' plane of the cosmic string spacetime (Eq.~(\ref{lp3})).

%\noindent 
For the Kerr-de Sitter universe, we have only retained terms linear in the rotation parameter $a$, which should be the case
far away from the source, where $\Lambda$ can play a prominant role.
 For the azimuthal photons, we have found non-vanishing contribution from $\Lambda$ (Eq.s~(\ref{lp19}),~(\ref{lp20}). 

%\noindent 
We note that for the two different spacetimes, $\Lambda$ contributes to the velocity shift in two very different qualitative ways. However, all of them are consistent with the polarization sum rule. Investigation of this effect for gravitons seems interesting, which we hope to address in a future work.

%%%%%%%%%%%%%%%%%%%%%%%%%%%%%%%%%%%%%%%%%%%%%%%%%%%%%%
\vskip 0.4cm
%\newpage
%\bigskip
\noindent
\textbf{Acknowledgement :}
This research is implemented under the ``ARISTEIA II" Action of the Operational Program ``Education and Lifelong Learning" and is co-funded by the European Social Fund (ESF) and Greek National Resources. 
%%%%%%%%%%%%%%%%%%%%%%%%%%%%%%%%%%%%%%%%%%%%%%%%%%%%%%%               REFS 
 
%\section{References}

\end{document}